\documentclass[seceq]{ptptex}
\usepackage{graphicx}
\usepackage{wrapft}
\usepackage{bm}% bold math

\def\itmb{\begin{itemize}}
\def\itme{\end{itemize}}
\def\enmb{\begin{enumerate}}
\def\enme{\end{enumerate}}
\def\eqnb{\begin{equation}}
\def\eqne{\end{equation}}
 
\def\PTP{Prog. Theor. Phys.(Kyoto)}

\def\PRL{Phys. Rev. Lett.}

\def\PRA{{Phys. Rev.} A}

\title{Entropy of Chaotic Currents in the Chua's Circuit\\
 and its HMM Analysis %\thanks{Grants or other notes
}
\author{%       %Use \scshape  for the family name
Sadataka \textsc{Furui}%
}

\inst{%     %Affiliation, neglected when [addenda] or [errata]
School of Science and Engineering, Teikyo University\\
 Utsunomiya 320-8551, Japan
}
\recdate{July 3, 2010}%            %Editorial Office will fill in this.
\abst{%
We study entropy of chaotic oscillation of electrical currents in the Chua's circuit controlled by triggering a pulse that brings the orbit that goes onto an unstable branch back to a stable branch.  A numerical simulation of the voltage of the two capacitors and the current that flows on an inductor of the Chua's circuit reveals various oscillation patterns as the conductance that is connected between the two capacitors and directly connected to an inductor is varied.
At small conductance, the Lissajous graph of the voltage of the two capacitors shows a spiral, while at high conductance a double scroll pattern appears. The entropy of the  current that flows on the inductor is alocal minimum in the spiral state which is in the steady state, while it is maximum in the stable double scroll state. 
The stable double scroll samples are analyzed by using the Hidden Markov Model (HMM) and the eigenvectors of the transition matrix of long time series are found to be strictly positive but those of unstable short time series have negative components.  We thus confirm maximum entropy production in the double scroll of the longest time series around the right fixed point, while the local minimum entropy production occurs in the spiral around the left fixed point.
}
\begin{document}
\newcommand{\ttbs}{\char'134}
\newcommand{\Slash}[1]{\ooalign{\hfil/\hfil\crcr$#1$}}
\maketitle
\section{Introduction}
%\label{intro}
In analyzing nature of complex systems  it is important to know whether the entropy of the system is maximum or minimum. We studied in \cite{FN06} the chaotic oscillation of the Chua's circuit\cite{CKM86}, which consists of two capacitors, inductor, two diodes, a variable resistance and an opeamp. The opeamp produces piecewise-linear but bending resistance, and thus the system is non-linear and when the variable resistance that is located between the two capacitors is decreased, the Lissajous graph of the voltage of the capacitor 1 and that of the capacitor 2 changes from a spiral to a double scroll. The oscillation becomes chaotic when the conductance of the variable resistance increases, but we showed in \cite{FN06} that the chaotic oscillation of currents can be controlled by triggering a pulse when the voltage of a capacitor in the Chua's circuit passes a certain voltage from the larger absolute value (right) to the lower absolute value (left). The pulse trigger violates the left-right symmetry, and the spiral occurs around left fixed point. The orbit returned from around right fixed point and which tend to move on the unstable branch is kicked by the pulse to one on a stable branch, and we could generate various oscillation patterns of the time series. 

We performed a simulation of the Chua's circuit and found that when conductance of the variable resister is larger than a critical value that changes the orbit from a spiral to a double scroll, but relatively smaller than another critical value from which a chaotic oscillation begins, there appears an orbit whose length of the time series is long. 
 But when the conductance becomes higher, transition to unstable orbit occurs and the period that the orbit is on the stable manifold becomes short.

In this paper, we analyze the entropy of the oscillation pattern of Chua's circuit, since whether the maximal entropy state and/or minimum entropy state can be detected is an important problem in the study general features of complex systems. 

Feynman said in the Lectures on Physics \cite{Fe65}, "If currents are made to go through a piece of material obeying Ohm's low, the currents distribute themselves inside the piece so that the rate at which energy is generated is as little as possible. Also we can say (if things are kept isothermal) that the rate at which energy is generated is a minimum." 

In the statistical physics, internal energy E, entropy S and the volume V satisfy the relation
\begin{equation}
 dE=T dS +p dV
\end{equation}
and at the constant pressure, $dS=\frac{1}{T}dE$. 

When an electrical circuit is attached to a battery and the heat $\mathcal Q$ and the work $\mathcal W$ are supplied to a system, the sum of the two makes the internal energy ${\mathcal U}$ and the heat divided by the absolute temperature $T$, $\displaystyle\frac{d\mathcal Q}{T}=dS$ is called the entropy change.
When local equilibrium exists in the system, the entropy change consists of that due to exchange of energy and/or matter with external system $dS_e$, and that due to irreversible processes in the system $dS_i$: $ dS=dS_e+dS_i $.

In the textbook on statistical physics of Landau and Lifshitz \cite{LL66} it is claimed that in the equilibrium states, there are stable and metastable states and that the stable state produces the local maximum of the entropy.  In non-equilibrium systems, Prigogine \cite{Pr01} showed that in the linearly interacting systems, the rate at which a stable, steady state, non-equilibrium system produces entropy internally is a minimum (minimum entropy production rule)\cite{JK03}. One could imagine that, when the temperature of subsystems is the same, minimum energy generation corresponds to the minimum entropy generation.

When there is a sudden temperature variation, the minimum entropy production rule even sufficiently close to the equilibrium is inadequate\cite{Landauer75}. 
The validity of entropy production principles for linear electrical circuits was discussed by Bruers et al.\cite{BMN07}. They started from the Langevin equations obtained by the Kirchhoff's laws with a Johnson-Nyquist noise at each dissipative element on linear electrical circuits and pointed out that the time-reversal odd physical component such as the current discussed in \cite{Landauer75} does not follow the minimum entropy production rule.

Dewar\cite{Dewar03} emphasized the constraint such as Kirchhoff's rule in the derivation of the entropy production rules. He derived the maximum entropy production by using the maximum path information entropy production method, and Bruers\cite{Bruers07} showed the relation between the equilibrium maximum entropy production in information system and non-equilibrium maximum entropy production in statistical systems. He showed how to derive minimum and maximum entropy production in statistical systems.

In the system of Chua's circuit, we consider the energy $E$ as  $VI=RI^2$ and $I^2$ is measured by using the Fourier transform and the Parcseval's formula. Due to Joule heating from the resistance and the pulse, the minimum entropy production rule could be violated, but from the Feynman's conjecture and Prigogine's theorem, the entropy of the stable spiral steady current is expected to be a local minimum.

In the case of capacitors, the entropy creation rate is \cite{Pr01}
\begin{eqnarray}
\frac{d_i S}{dt}&=&\frac{V_c I}{T}=\frac{V c}{T}\frac{dQ}{dt}=-\frac{C}{T}V c\frac{dV c}{dt}\nonumber\\
&=&-\frac{1}{T}\frac{d}{dt}(\frac{C{V c}^2}{2})=-\frac{1}{T}\frac{d}{dt}(\frac{Q^2}{2C})>0.
\end{eqnarray}

In the case of inductor,
\begin{equation}
\frac{d_i S}{dt}=-\frac{1}{T}\frac{d}{dt}(\frac{LI^2}{2})=-\frac{LI}{T}\frac{dI}{dt}=\frac{V LI}{T}>0.
\end{equation}

Since we trigger a pulse when the orbit tends to escape from the stable manifolds, we consider ensembles of different time length.  Entropy production and escape rates was discussed in \cite{MN06}. They start from stationary ergodic Markov process with the state $X_t, t\geq 0$ at time $t$.  They fixed $\tau>0$ and the average of $X_\tau, X_{2\tau},\cdots, X_T$ is defined as the time average.
The states of the Markov process are defined as intervals $[A/\epsilon, B/\epsilon] \in {\bf R}$, where $\epsilon$ defines a scale such that $x=\epsilon X\in [0,1]$ and the probability that the system has a value $\epsilon X=x$ at time $t$ is defined as $p(x,t)$.  In the stationary state $p(x)$ is assumed to be proportional to $e^{-S(x)/\epsilon}$, where $S(x)$ is the entropy function.  They showed a relation between the entropy function and the Hamilton's principal function.

Orbits of Markov process on stable and unstable manifolds are studied in the system of hyperbolic differential equation\cite{Bowen75}.  When an orbit is close to the fixed point defined by the process, Perron-Frobenius theorem\cite{Se06} asserts that the transition matrix has a positive eigenvalue and its eigenvectors have all positive components. When there are intersections of stable and unstable manifolds, the theorem is modified by Ruell (Ruelle's Perron-Frobenius theorem) such that theorem applies when the sequence is long enough \cite{Bowen75}.
Since tangencies between the stable and unstable manifolds occur frequently\cite{Newhouse79}, the theorem could depend on the nature of the fixed point.

In speech recognitions and pattern recognitions of time series's data, the Hidden Markov Model (HMM) which is an application of the Bayes statistical theory is well known. In this method, one tries to determine hidden parameters from observed parameters. The extracted model parameters are used to perform further analysis and parameters that define the time series of the Markov Process in the equilibrium are derived by iteration \cite{Ba70,We03}. Application of the HMM to double scroll time series of Chua's circuit was tried in 90's \cite{DF93} but not successfully, since they could not predict to which fixed point the orbit will approach in the future. Since we trigger a pulse when a orbit around right fixed point passes to unstable branch, the unstable orbits around right fixed point are suppressed, and one can obtain parameters of stable orbits around right fixed point.

 We apply the HMM in the analysis of the time series of the voltage of the capacitor 1 ($v_{C1}$), that of the capacitor 2 ($v_{C2}$)  and the current on the inductor ($i_L$). We record the  data sets of $\{v_{C1}, v_{C2}, i_L\}$, and consider the Poincar\'e surface specified by $i_L=\pm GF$, where $G$ is the conductance and $F$ is defined from the fixed point of the double scroll orbits $(\bar v_{C1}, \bar v_{C2})=(\pm F,0)$.  We define the distance between the point at $t_i$ on the orbit and the fixed point on the Poincar\'e surface $x(t_i)=\sqrt{(F-|v_{C1}|)^2+v_{C2}^2}$ where $t_i$ is the time when the orbit crosses the Poincar\'e surface. We also measure the binary (0,1) sequence $b_1, b_2,\cdots$ whether the sign of $v_{C1}$ is positive or negative between $t_i$ and $t_{i+1}$.

 Since the HMM measures the forward time series and backward time series towards fixed points separately, we study the dependence of the Perron-Frobenius theorem on the fixed points of the double scroll and correlation to the length of the time series between the pulses triggered as the orbit passes a plane in the phase space. 

The contents of this paper are as follows.
In sect.2, we present the differential equation used in analyzing the Chua's circuit and in sect.3, the result of entropy analysis. In sect.4, the HMM applied to the double scroll system is explained and in sect.5, the result of HMM is summarized. Discussion and conclusion are given in sect.6.

\section{The differential equation of the Chua's circuit}
Chua's circuit consists of an autonomous circuit which contains three-segment piecewise-linear resistor, two capacitors, one inductor and a variable resistor.

The equation of the circuit is described by
\begin{equation}\label{chua_bare}
\left\{\begin{array}{l}
C_1\frac{d}{dt}v_{C1}=G(v_{C2}-v_{C1})-\tilde g(v_{C1},m_0,m_1)\\
C_2\frac{d}{dt}v_{C2}=G(v_{C1}-v_{C2})+i_L\\
L\frac{d}{dt}i_L=-v_{C2}
\end{array}\right.
\end{equation}
where $v_{C1}$ and $v_{C2}$ are the voltages of the two capacitors (in V), $i_L$ is the current that flows in the inductor (in A), $C_1$ and $C_2$ are capacitance (in F), $L$ is inductance (in H) and $G$ is the conductance of the variable resistor (in $\Omega^{-1}$). The three-segment piecewise-linear resistor which constitutes the non-linear element is characterized by,
\[
\tilde g(v_{C1},m 0,m 1)=(m_1-m_0)(|v_{C1}+B_p|-|v_{C1}-B_p|)/2+m_0 v_{C1}, 
\]
where $B_p$ is chosen to be 1V, $m_0$ is the slope (mA/V) outside $|v_{C1}|>B_p$  and $m_1$ is the slope inside $-B_p<v_{C1}<B_p$. 
The function $\tilde g(v_{C1},m_0,m_1)$ can be regarded as an active resistor. 

\begin{figure}[htb]
\begin{center}
\includegraphics[width=12cm,angle=0,clip]{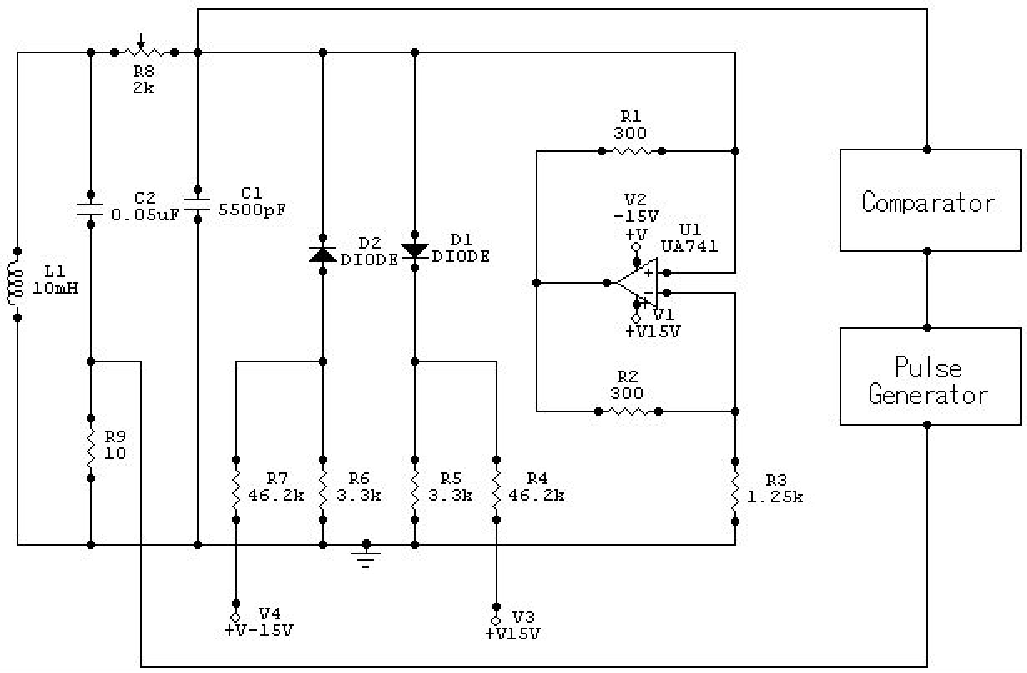}
\end{center}
\caption{Chua's circuit with a pulse generator used in the experiment of \cite{FN06}.}\label{chua_circuit}
\end{figure}

As the conductance between the two capacitors is increased, the spiral changes to double scroll. Details of the Chua's circuit with impulse generator and its bifurcation analysis are shown in \cite{FN06} and the network is shown in Fig.\ref{chua_circuit}.

Experimental data of the transition from spiral to double scroll as a function of the variable resistance (R8 in the Fig.\ref{chua_circuit}) are shown in Appendix A1.  

\begin{figure}[htb]
\begin{minipage}[b]{0.47\linewidth}
\begin{center}
\includegraphics[width=6cm,angle=0,clip]{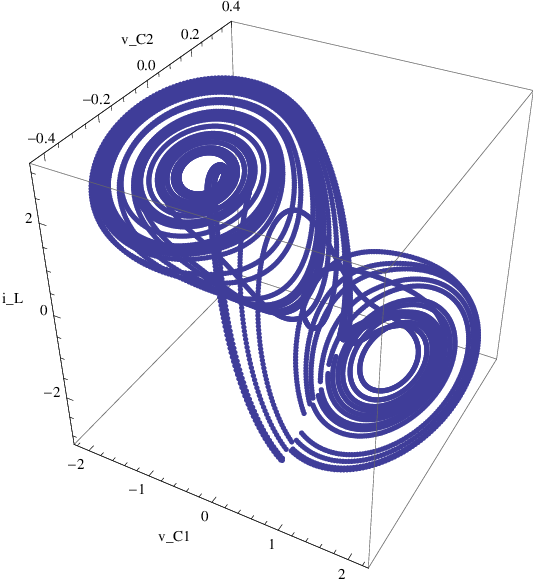}
\end{center}
\caption{An example of the orbit in $\{v_{C1},v_{C2},i_L\}$ space. ($G=0.7052$)}\label{fig:1}\end{minipage}
\hfill
\begin{minipage}[b]{0.47\linewidth}
\begin{center}
\includegraphics[width=6cm,angle=0,clip]{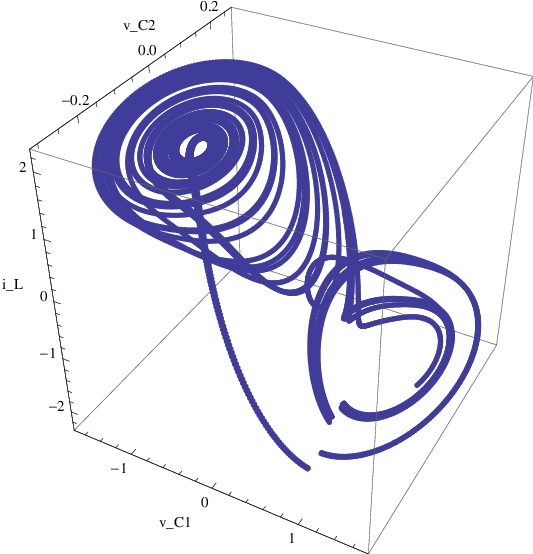}
\caption{An example of the orbit in $\{v_{C1},v_{C2},i_L\}$ space. ($G=0.7429$)}
\label{fig:2}
\end{center}
\end{minipage}
\end{figure}

Simulation data of entropy production from the capacitors and the inductor are shown in Appendix A2. 
As examples of the orbit, we take  $G=0.7429$ which is close to the transition point of spiral to double scroll, and $G=0.7052$ which is in the double scroll area. 
The orbits in $\{v_{C1},v_{C2},i_L\}$ space are shown in Fig.\ref{fig:1} and in Fig.\ref{fig:2} respectively for $G=0.7052$ and for $G=0.7429$.

\section{The entropy analysis of the electrical current}

For an oscillation of a period $\mathcal T$, the expansion of a time series's data $f(t)$ is given as 
\[
f(t)=\frac{a_0}{2}+\sum_{k=1}^\infty (a_k\cos 2\pi k t+b_k\sin 2\pi k t)
\]
\[
a_k=\frac{2}{\mathcal T}\int_0^{\mathcal T} f(t)\cos 2\pi k t dt
\]
\[
b_k=\frac{2}{\mathcal T}\int_0^{\mathcal T} f(t)\sin 2\pi k t dt
\]
and by applying the Parseval's formula, we obtain
\begin{equation}
\int_0^{\mathcal T} f^2(t)dt={\mathcal T} \frac{a_0^2}{4}+\frac{\mathcal T}{2}\sum_{k=1}^\infty (a_k^2+b_k^2)
\end{equation}
The integral over a period $\mathcal T$ becomes $\displaystyle \int_0^T \frac{L}{2}I(t)^2 dt$,
$\displaystyle \int_0^{\mathcal T} \frac{C}{2}V(t)^2 dt$, $\displaystyle \int_0^{\mathcal T} \frac{R}{2}I(t)^2dt$. 
We plot the zero-mode subtracted average 

$\displaystyle S_C=\frac{1}{\mathcal T}\int_0^{\mathcal T} V(t)^2-\frac{a_0^2}{4}=\frac{1}{2}\sum_{k=1}^\infty (a_k^2+b_k^2)$, etc.

\begin{figure}[htb]
\begin{minipage}[b]{0.47\linewidth}
\begin{center}
\includegraphics[width=6cm,angle=0,clip]{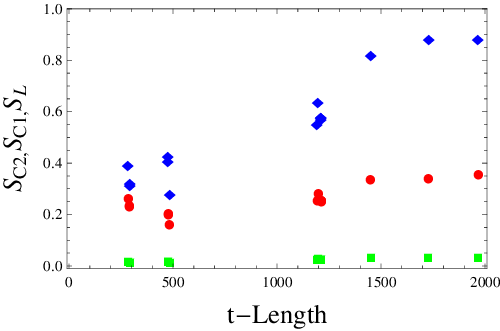}
\caption{The zero-mode subtracted entropy of the $G=0.7052$ system as a function of the length of the time series. The diamond is the entropy of the current $S_L$, the disk is that of the capacitor 1 $S_{C1}$ and square is that of the capacitor 2 $S_{C2}$.  The orbits on $W^u(R)$ that rotates around the right fixed point shows a local maximum of the entropy.}\label{fig:3}
\end{center}
\end{minipage}
\hfill
\begin{minipage}[b]{0.47\linewidth}
\begin{center}
\includegraphics[width=6cm,angle=0,clip]{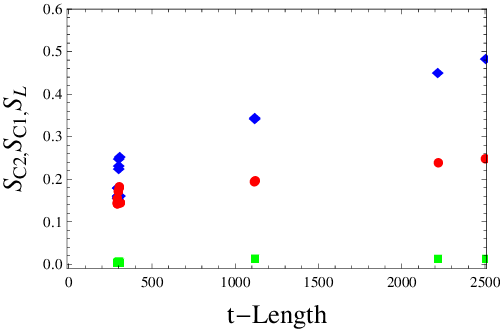}
\caption{The zero-mode subtracted entropy of the $G=0.7429$ system as a function of the length of the time series. The meanings of the symbols are the same as Fig.\ref{fig:3}. }\label{fig:4}
\end{center}
\end{minipage}
\end{figure}

We measure the average entropy of the electrical current between the periods when  the pulses are triggered. The zero-mode subtracted entropy as a function of the length of the period of $G=0.7052$ is shown in Fig. \ref{fig:3}, and that of $G=0.7429$ is shown in Fig.\ref{fig:4}. We define the orbits around the left fixed point as $W^s(L)$ and the orbits around the right fixed point as $W^u(R)$. The orbits start from around the left fixed point.

\begin{figure}[htb]
\begin{minipage}[b]{0.47\linewidth}
\begin{center}
\includegraphics[width=6cm,angle=0,clip]{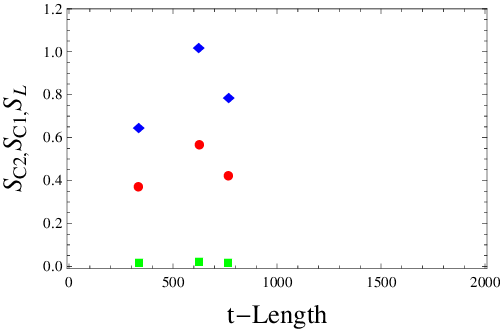}
\caption{The zero-mode subtracted entropy of an orbit on $W^s(L)$ of the $G=0.7052$ system as a function of the length of the time series. }\label{fig:5}
\end{center}
\end{minipage}
\hfill
\begin{minipage}[b]{0.47\linewidth}
\begin{center}
\includegraphics[width=6cm,angle=0,clip]{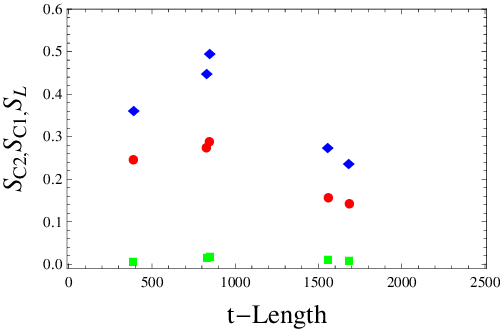}
\caption{The zero-mode subtracted entropy of an orbit on $W^s(L)$ of the $G=0.7429$ system as a function of the length of the time series.}
\label{fig:6}\end{center}
\end{minipage}
\end{figure}

\begin{figure}[htb]
\begin{minipage}[b]{0.47\linewidth}
\begin{center}
\includegraphics[width=6cm,angle=0,clip]{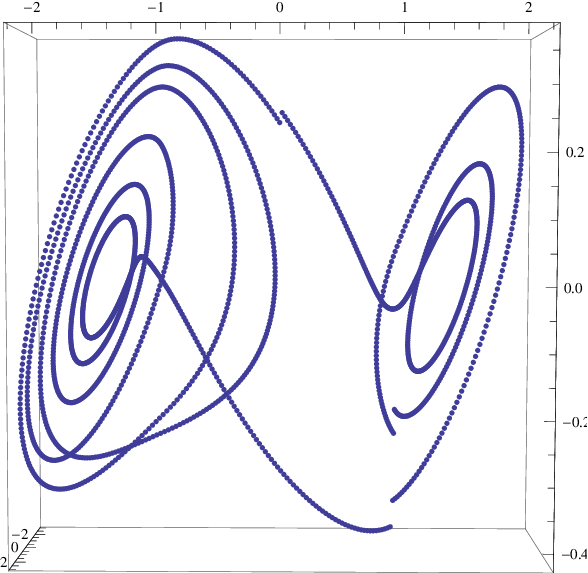}
\end{center}
\caption{The orbit on $W^u(R)$ and $W^s(L)$ of the $G=0.7052$ system.}
\label{fig:7}
\end{minipage}
\hfill
\begin{minipage}[b]{0.47\linewidth}
\begin{center}
\includegraphics[width=6cm,angle=0,clip]{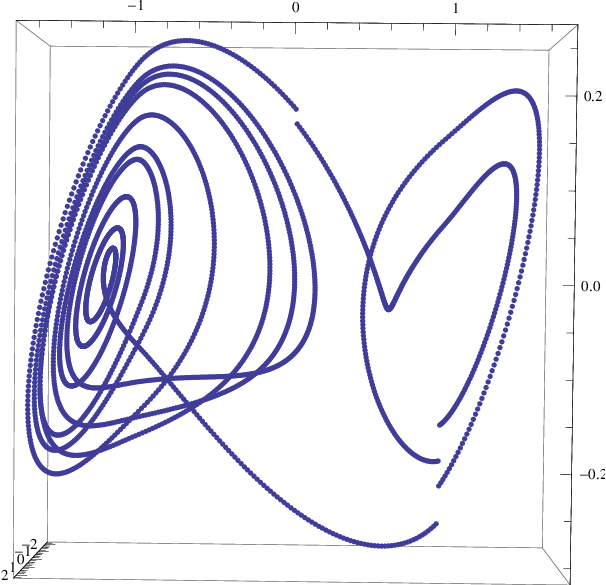}
\end{center}
\caption{The orbit on $W^u(R)$ and $W^s(L)$ of the $G=0.7429$ system. }
\label{fig:8}
\end{minipage}
\end{figure}

The zero-mode contribution to $S_L$ is larger in the case of $G=0.7429$ than in the case of $G=0.7052$. When the length of the double scroll is the longest, its entropy becomes maximum, which suggests Ruell's Perron Frobenius theorem.  
In order to extract parameters that characterize the oscillation patterns, we assemble samples whose period ${\mathcal T}$ is longer than 1000. An example of the ordinary orbit of $G=0.7052$ is shown in Fig.\ref{fig:7} and that of $G=0.7429$ is Fig.\ref{fig:8}.  .

\begin{figure}[htb]
\begin{minipage}[b]{0.47\linewidth}
\begin{center}
\includegraphics[width=6cm,angle=0,clip]{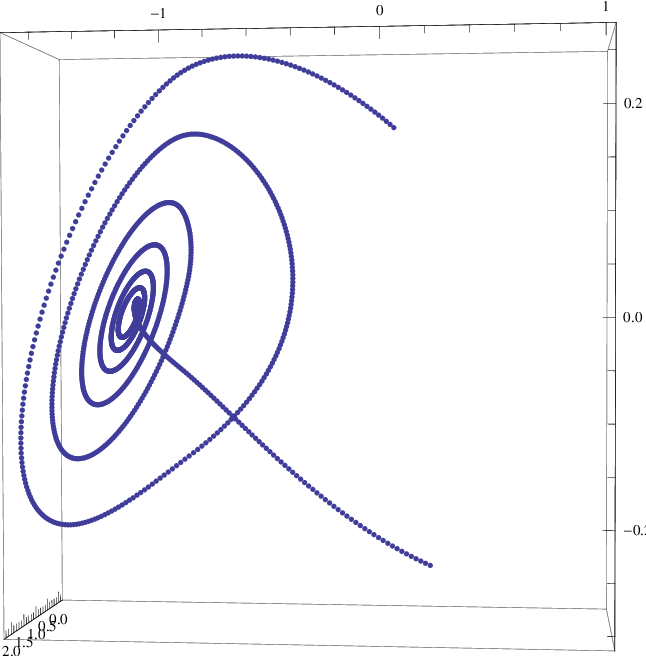}
\end{center}
\caption{The long orbit on $W^s(L)$ of the $G=0.7429$ system(4c). }
\label{fig:9}
\end{minipage}
\hfill
\begin{minipage}[b]{0.47\linewidth}
\begin{center}
\includegraphics[width=6cm,angle=0,clip]{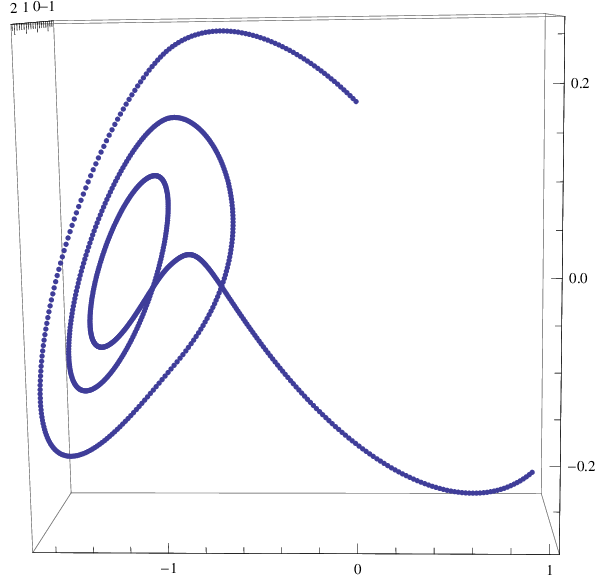}
\end{center}
\caption{The short orbit on $W^s(L)$ of the $G=0.7429$ system(1c).}
\label{fig:10}
\end{minipage}
\end{figure}

In addition to the orbit whose entropy increases monotonically, there is a branch whose entropy reaches a maximum and then decreases as shown in Fig.\ref{fig:5} and in Fig.\ref{fig:6}. The decrease of the entropy means the orbit approaches  a fixed point whose entropy is local minimum.
These two branches are separable in the case of $G=0.7052$ as Fig.\ref{fig:3} and Fig.\ref{fig:5}.  When $G=0.7429$, two orbits overlap as Fig.\ref{fig:4} and F\ref{fig:6} show. The typical double scroll orbit of $G=0.7052$ in Fig.\ref{fig:7} and that of $G=0.7492$ in Fig.\ref{fig:8} indicate that long circulation around right fixed point appears freqently in $G=0.7052$ but not in $G=0.7492$.  Fig.\ref{fig:9} and Fig.\ref{fig:10} are examples of long and short orbits in $G=0.7492$.

\section{The HMM analysis of the electrical current}
In general, a sequence of events is a Markov process when the time series is defined by the state of one step before. When one restricts chaotic oscillation of the voltage of capacitors to the stable double scroll, and by taking into account the pulse that kicks the current on the unstable orbit to a stable orbit of the double scroll, one could make the system satisfy the detailed balance, and  
 the time series of the stable double scroll can be regarded as a Markov chain.  In this paper, we restrict the analysis of samples whose period $\mathcal T$ is longer than 1000.  All the orbits of this sample are assumed to be absorbed in a limit orbit.

 We identify the position of the orbit at time $t$ following Hayes, Grebogi and Ott\cite{HGO93}, in which the distance of the position on the orbit from the fixed point is calculated as
\begin{equation}
x(t_j)=\sqrt{(F-|v_{C1}(t_j)|)^2+v_{C2}(t_j)^2 }
\end{equation}
and specify around which point it circulates by a coding function $r(x)$ which is defined as follows.

 We define the binary time series $b_1,b_2,b_3,\cdots$ assigned at each time when the orbit crosses the Poincar\'e surface of $i_L=\pm GF$, according to the voltage of the capacitor $v_{C1}$ is positive(1) or negative(0).  From the data of $b_n$, we define 
\begin{equation}
r=\sum_{n=1}^\infty b_n 2^{-n}.
\end{equation}
The symbols that occur in earlier times are given greater weight.

Using the list $R=r(1),r(2),\cdots,r(\mathcal T)$ and the list $X=x_1,x_2,\cdots, x_{\mathcal T}$,  we denote by $w_i$ the pair of list $(r(t),x(t))$  and the probability that the system is in this state as
\begin{equation}
P(w_i|S)=\frac{P(S|w_i)P(w_i)}{P(S)}.
\end{equation}

We call the state at $t=t_1$ when the orbit passes the Poincar\'e surface and the measurement of the oscillation states as $i$ and the state at $t=t_2$  when the orbit passes the Poincar\'e surface at the next time as $j$.
The transition matrices $a_{ij}$ for arbitrary $j$ can be derived from the sample average when samples of the similar length $\mathcal T$ are assembled.  

We define a model $M$ that the system makes a transition from a state 1 to a state 2 and output the data $x_1$, then makes a transition from 2 to a state 3 as giving an output $x_2$, and then output $x_3$, and the probability
\begin{equation}
P(X,S|M)=a_{12}b_2(x_1)a_{22}b_2(x_2)a_{23}b_3(x_3)\cdots.
\end{equation}
We try to maximize the probability
\begin{equation}
P(X|M)=\sum_R a_{r(0)r(1)}\prod^{\mathcal T}_{t=1}b_{r(t)}(x_t)a_{r(t)r(t+1)}.
\end{equation}

We define the distribution of the observation symbol $b_j(k)$ as the probability that the system in the state $j$ gives the output $y_k$, which is calculated by integrating the distribution given by a list vector $b_j(x_t)$. 

\subsection{Forward probability}
When the number of states of the model M is N, the forward probability 
\begin{equation}
\alpha_t(j)=P(x_1,\cdots,x_t,r(t)=j|M)
\end{equation}
can be evaluated from the recursion
\begin{equation}
\alpha_{t+1}(j)=[\sum_{i=1}^N \alpha_t(i)a_{ij}]b_j(x_{t+1})
\end{equation}

The initial condition is such that the probability that the system is in the state $i$ is $\pi_i$: $\alpha_1(i)=\pi_i b_i(x_1)$.
In the calculation $\alpha_{t+1}$, $b_j(x_t)$ is a sum of the components of a list vector given from the distribution of $x$ at time $t$. 
 
The $a_{ij}$ and $b_j(k)$ are updated in the backward probability calculation.

\subsection{Backward probability}
The backward probability $\beta_t(j)$ is defined as
\begin{equation}
\beta_t(j)=P(y_{t+1},\cdots,y_{\mathcal T}|s(t)=j,M)
\end{equation}
The initial condition $\beta_{\mathcal T}(i)=b_{\mathcal T}(i)$ is defined from the sample average at $t=\mathcal T$.
$\beta_t(i)$ is calculated by the recursion
\begin{equation}
\beta_t(i)=\sum_{j=1}^{N}a_{ij}b_j(x_{t+1})\beta_{t+1}(j),
\end{equation}
where $b_j(x_{t+1})$ is a list vector, and $\beta_{t-1}(i)$ is the sum of the components of the list vector.

When the model $M$ is specified by $\lambda=(R,X,\pi)$, the probability the data $x$ is output is 
\begin{equation}
Pr(x|R,X,\pi)=\sum_{i=1}^N\sum_{i=1}^N\alpha_t(i)a_{ij}b_j(x_{t+1})\beta_{t+1}(j)
\end{equation}
We define
\begin{equation}
\xi_t(i,j)=\frac{\alpha_t(i)a_{ij}b_j(x_{t+1})\beta_{t+1}(j)}{Pr(x|R,X,\pi)},
\end{equation}
and $\gamma_t(i)=\sum_{j=1}^N\xi_t(i,j)$. The probability $Pr(x|R,X,\pi)$ is a sum of the probabilities such that the components of the list vector $b_j(x_{t+1})$  are summed up.

The transition matrix $a_{ij}$ is updated as
\begin{equation}
a_{ij}=\frac{\sum_{t=1}^{{\mathcal T}-1}\xi_t(i,j)}{\sum_{t=1}^{{\mathcal T}-1}\gamma_t(i)}
\end{equation}

The distribution $b_j(k)$ is updated as
\begin{equation}
b_j(k)=\frac{\sum_{t=1,y_t=k}^{{\mathcal T}-1}\xi_t(i,j)}{\sum_{t=1}^{{\mathcal T}-1}\gamma_t(i)}
\end{equation}

\section{The results of the HMM analysis}

Among samples with the same conductance but different initial conditions, we pick up about 10 samples of the data of $G=0.7052$, whose length of the period $\mathcal T$ is larger than 1000. 

The states of the time series are assigned by $r$ which runs from 0(state 1), 1/32(state 2), 3/32(state 3),  7/32(state 4), 15/32(state 5), 1/2(state 6), 31/32(state  7) and the final state (state 8). 
Since the initial state has $r=\frac{1}{2}$, the first state is assigned as the 6th state, and since the next state has $r=0$, the second state is assigned as the 1st state. 

The state in the 6th state is transferred to the 1st state with the probability of 100\%. Then in the next step it stays in the 1st state with the probability of 96\% and goes to the 2nd state with the probability of 4\%. This kind of information is contained in the matrix $a_{ij}$.

The transition matrix $a_{ij}$  obtained from the long $\mathcal T$ samples is shown in Table \ref{tab:1}.
\begin{table}[htb]
\begin{tabular}{lllllllll}
i$\setminus$ j & 1 & 2 & 3 & 4 & 5 & 6 & 7 & 8 \\
\hline\noalign{\smallskip}
1&0.955882 &  0.0441176 &  0. &  0. &  0. &  0. &  0. &  0. \\   
2& 0. &  0. &  0.857143 &  0. &  0. &  0. &  0. &  0.142857  \\  
3& 0. &  0. &  0. &  1. &  0. &  0. &  0. &  0. \\   
4& 0. &  0. &  0. &  0. &  1. &  0. &  0. &  0. \\   
5& 0. &  0. &  0. &  0. &  0. &  0. &  1. &  0. \\   
6& 1. &  0. &  0. &  0. &  0. &  0. &  0. &  0. \\   
7& 0. &  0. &  0. &  0. &  0. &  0. &  0.4 &   0.6 \\
8& 0. &  0. &  0. &  0. &  0. &  0. &  0. &  1.  \\
\noalign{\smallskip}\hline
\end{tabular}
\caption{The $a_{ij}$ matrix of the $G=0.7052$ long $\mathcal T$ samples.}\label{tab:1}
\end{table}

When the samples are restricted to short $\mathcal T$ samples, we obtained the  $a_{ij}$ matrix is shown in Table \ref{tab:2}.
\begin{table}[htb]
\begin{tabular}{lllllllll}
i$\setminus$ j & 1 & 2 & 3 & 4 & 5 & 6 & 7 & 8 \\
\hline\noalign{\smallskip}
1&0.849673& 0.150327& 0.& 0.& 0.& 0.& 0.& 0.\\
2&0.& 0.& 0.913043& 0.&  0.& 0.& 0.& 0.0869565\\ 
3&0.& 0.& 0.& 0.952381& 0.& 0.& 0.&   0.047619\\
4&0.& 0.& 0.& 0.& 0.95& 0.& 0.& 0.05\\ 
5&0.& 0.& 0.& 0.&   0.& 0.& 0.947368& 0.0526316\\ 
6&1.& 0.& 0.& 0.& 0.& 0.& 0.& 0.\\ 
7&0.&   0.& 0.& 0.& 0.& 0.& 0.217391& 0.782609\\ 
8&0.& 0.& 0.& 0.& 0.& 0.&   0.& 1.\\
\noalign{\smallskip}\hline
\end{tabular}
\caption{The $a_{ij}$ matrix of the $G=0.7052$ short $\mathcal T$ samples.}\label{tab:2}
\end{table}

We define the probability $b_t(i)$ such that when the orbit passes a point $x_t$ on the Poincar\'e surfaces at the time 't', the state is in the 'i'. 
The sequence dependent state probability $b_t(i)$ of $G=0.7052$, long $\mathcal T$ samples are given in Table \ref{tab:3}.
The corresponding data of short $\mathcal T$ samples are given in Table \ref{tab:4}.
\begin{table}[htb]\begin{tabular}{lllllllll}
t$\setminus$ i & 1 & 2 & 3 & 4 & 5 & 6 & 7 & 8 \\
\hline\noalign{\smallskip}
1 & 0.& 0.& 0.& 0.& 0.& 1.& 0.& 0.\\
2 & 1.& 0.& 0.& 0.& 0.& 0.& 0.& 0.\\ 
3 & 1.& 0.& 0.& 0.& 0.& 0.& 0.& 0.\\ 
4 & 1.& 0.& 0.& 0.& 0.& 0.& 0.& 0.\\
5 & 1.& 0.& 0.& 0.& 0.& 0.& 0.& 0.\\
6 & 1.& 0.& 0.& 0.& 0.& 0.& 0.&  0.\\ 
7 & 1.& 0.& 0.& 0.& 0.& 0.& 0.& 0.\\
8 & 1.& 0.& 0.& 0.& 0.& 0.& 0.& 0.\\ 
9 & 1.& 0.& 0.& 0.& 0.& 0.& 0.& 0.\\
10 & 1.& 0.& 0.& 0.& 0.& 0.& 0.& 0.\\
11 & 1.& 0.& 0.& 0.& 0.& 0.& 0.& 0.\\ 
12 & 1.& 0.& 0.& 0.& 0.& 0.& 0.& 0.\\
13 & 1.& 0.& 0.& 0.& 0.& 0.& 0.& 0.\\
14 & 1.& 0.& 0.& 0.& 0.& 0.& 0.&  0.\\
15 & 1.& 0.& 0.& 0.& 0.& 0.& 0.& 0.\\
16 & 1.& 0.& 0.& 0.& 0.& 0.& 0.&  0.\\ 
17 & 1.& 0.& 0.& 0.& 0.& 0.& 0.& 0.\\
18 & 0.8& 0.2& 0.& 0.& 0.& 0.& 0.& 0.\\ 
19 & 0.8& 0.& 0.2& 0.& 0.& 0.& 0.& 0.\\
20 & 0.7& 0.1& 0.& 0.2& 0.&  0.& 0.& 0.\\ 
21 & 0.7& 0.& 0.1& 0.& 0.2& 0.& 0.& 0.\\ 
22 & 0.4& 0.3& 0.&  0.1& 0.& 0.& 0.2& 0.\\ 
23 & 0.4& 0.& 0.3& 0.& 0.1& 0.& 0.2& 0.\\
24 & 0.3& 0.1& 0.& 0.3& 0.& 0.& 0.1& 0.2\\
25 & 0.3& 0.& 0.& 0.& 0.3& 0.& 0.& 0.4\\
\noalign{\smallskip}\hline
\end{tabular}
\caption{The sequence dependent state probability $b_t(i)$ of $G=0.7052$ long samples.}\label{tab:3}
\end{table}

\begin{table}[htb]
\begin{tabular}{lllllllll}
t$\setminus$ i & 1 & 2 & 3 & 4 & 5 & 6 & 7 & 8 \\
\hline\noalign{\smallskip}
1 &0. & 0. & 0. & 0. & 0. & 1. & 0. & 0. \\
2 &1. & 0. & 0. & 0. & 0. & 0. & 0. & 0. \\
3 &1. & 0. & 0. & 0. & 0. & 0. & 0. & 0. \\
4 &1. & 0. & 0. & 0. & 0. & 0. & 0. & 0. \\
5 &1. & 0. & 0. & 0. & 0. & 0. & 0. & 0. \\
6 &0.142857 & 0.857143 & 0. & 0. & 0. & 0. & 0. & 0. \\
7 &0.142857 & 0. & 0.857143 & 0. & 0. & 0. & 0. & 0. \\
8 &0.0357143 & 0.107143 & 0. & 0.857143 & 0. & 0. & 0. & 0. \\
9 &0.0357143 & 0. & 0.107143 & 0. & 0.857143 &  0. & 0. & 0. \\
10 &0. &  0.0357143 & 0. & 0.107143 & 0. & 0. & 0.857143 & 0. \\
11 &0. & 0. & 0. & 0. & 0.107143 & 0. & 0.0714286 & 0.821429 \\
12 &0. & 0. & 0. & 0. & 0. & 0. & 0. & 1.\\
13 &0. & 0. & 0. & 0. & 0. & 0. & 0. & 1.\\
14 &0. & 0. & 0. & 0. & 0. & 0. & 0. & 1.\\
\noalign{\smallskip}\hline
\end{tabular}
\caption{The sequence dependent state probability $b_t(i)$ of $G=0.7052$ short samples.}\label{tab:4}
\end{table}

In the HMM, the values of $a_{ij}$ and the $b_t(i)$ in the equilibrium are obtained by iteration. 
 The Perron-Frobenius theorem says that a dynamical system with a finite number of states can be guaranteed to converge to an equilibrium distribution $\rho^*$, if the computer algorithm is Markovian, is ergodic and satisfies detailed balance\cite{Se06}.  

Before iteration, the $a_{ij}$ of the long $\mathcal T$ samples has three non-zero eigenvalues which are 1, 0.96 and 0.4.
After one iteration $a_{12}=0.044$ and $a_{78}=0.6$ changes to 0 and nonzero eigenvalues become 1,1 and 0.18. Although an eigenvector of $a_{ij}$ before iteration has negative components, after an iteration, all the eigenvectors have positive components. It is a general property that a Markov chain should satisfy due to the Perron-Frobenius theorem\cite{Se06}.

In the case of short $\mathcal T$ samples, non-zero eigenvalues before iteration are 1, 0.850 and 0.217. Main difference from the long $\mathcal T$ samples is that the sign of two components of an eigenvector becomes negative, which suggests that the system is unstable.  

The qualitative difference of $G=0.7052$ and $G=0.7492$ can be understood in the Markov partition. 
 A rectangle $R$ in the phase space of the orbits is covered by proper rectangles that satisfy for $x$ in a rectangle $R$ and the diffeomorphic map $f$,
\begin{itemize}
\item (a) ${\rm int}(R_i)\cap {\rm int} (R_j)=\phi$
\item (b) $fW^u(x,R_i)\supset W^u(fx,R_j)\, {\rm and }\\
fW^s(x,R_i)\subset W^s(fx,R_j)\, {\rm when }\, x\, \in\, {\rm int}(R_i), fx \,\in\, {\rm int}(R_j)$.
\end{itemize}

In the covering of the rectangle $\{T_1,T_2,\cdots,T_r\}$, $T_j\cap T_k\ne \phi$, one could separate 

$T^2_{j,k}=\{x\in T_j:W^u(x,T_j)\cap T_k\ne\phi, W^s(x,T_j)\cap T_k=\phi\}$ and 

$T^3_{j,k}=\{x\in T_j:W^u(x,T_j)\cap T_k=\phi, W^s(x,T_j)\cap T_k\ne\phi \}$\\
in our discretized time series's data of $G=0.7052$. But we had the impression that the wild hyperbolic sets \cite{Newhouse79} i.e. tangency of unstable and stable manifolds appear quite often in $G=0.7492$.

\section{Discussion and Conclusion}
We studied the entropy of the electrical current and the parameters of the HMM of the Chua's circuit. We observed that the entropy of the steady spiral is a local minimum and that of the stable double scroll is a local maximum and the sample that has longest period ${\mathcal T}$ has the maximum entropy. There appears samples, which change their double scroll orbits to spiral-like orbits which make also a local maximum of the entropy. Restricting samples to those of the stable double scroll which have the period $\mathcal T$ longer than 1000 in the case of $G=0.7052$, we could make the time series satisfy the Markov chain condition and obtain the transition matrix $a_{ij}$ whose largest eigen value is 1 and the components of the eigenvector are positive.

The Perron-Frobenius theorem \cite{CR99,Bo09} says that for a square nonnegative matrix some power of which is positive, there is a simple root $\lambda$ of the characteristic polynomial which is strictly greater than the modulus of any other roots, and  $\lambda$ has strictly positive eigenvectors. In the proof of this theorem, Brouwer's fixed point theorem is used. 

In the short $\mathcal T$ samples of the time series of $G=0.7052$, flows to the equilibrium stable spiral state and non-equiliblium stable double scroll occurs, and we find that the eigenvectors of $a_{ij}$ of short $\mathcal T$ sample is not strictly positive.
The topology of the underlying dynamical system invalidates the application of the Brower's fixed point theorem in this case and introduction of branched manifold and Markov decomposition of the manifold\cite{Bowen75,CR99} is necessary.  We think the long $\mathcal T$ samples of $G=0.7052$ are in the near-equilibrium linear response regime\cite{Bruers07} and the orbit in the asymptotic state is in $W^u(R)$ while the asymptototic states of samples of $G=0.7429$ are in $W^s(L)$ but the tangencial transition to the manifold with $W^u(L)$ is frequent.

Although we studied a specific dynamical system of electrical circuits, it is the first successful application of HMM to the chaotic double scroll system. There are interesting problems generic to non-equilibrium dynamical systems. We confirm maximum entropy production in the double scroll orbits in the near-equilibrium linear response regime, while the entropy production in the spiral is found to be a local minimum.

Properties of the fixed point is important also in a non-linear system like QCD. Simulation analyses suggest that there is an ultraviolet fixed point and an infrared fixed point. Gauge invariance of the ultraviolet fixed point is established, but the structure of the infrared fixed point is complicated due to the Gribov copies and relaively large color anti-symmetric part as compared to the color symmetric part in the ghost propagator\cite{SF08}. Among many orbits tending to a local minimum, one tries to find a unique gauge. If the analogies of the ultraviolet fixed point and the right fixed point of the double scroll, and the infrared fixed points and the left fixed points of the double sroll work, further analyses of the Chua's circuit will allow to find a way to pick up an orbit that flows to a unique gauge when there are many local minima, or will assert its difficulty. 

%\vskip 0.5 4true cm
\section*{Acknowledgements}
 The author thanks  Mr. N. Hoshi for preparing figures and tables and Dr. K. Moriya for helpful discussions. The author acknowledges useful information on entropy production rules from anonymous one of two referees of a journal.
\newpage

\section*{Appendix} %Empty argument \section{} yields `Appendix'. 

In this appendix, we show an example of the transition from a spiral to double scroll observed experimentally on the Chua's circuit\cite{FN06}, and the entropy of the capacitors 1, 2 and the inductor measured by the simulation\cite{NH10}. 
\vskip 0.2 true cm
\leftline{{\bf A1.} The transition from a spiral to double scroll}
\vskip 0.2 true cm
When conductance is small, the transition from an orbit around one fixed point to an orbit around another fixed point does not occur and the spairal pattern appears as shown in Table \ref{tab:5}. The number of cycle increases as the conductance increases and transition to double scroll occur. The number of circles around left fixed point and around the right fixed point becomes unequal at some conductance. The symbol $L3R4$ means 3 circulations around the left fixed point and 4 circulation around right fixed point. The experimental data suggest that a long orbit that circulates around the right fixed point occurs when the conductance is small. The simulation data of $G=0.7052$ corresponds to this case.

\begin{table}[htb]
\begin{tabular}{llll}
 type& resistance ($R[k\Omega]$)& No. of periods& conductance ($1/R$)\\
\hline\noalign{\smallskip}
spiral & 1.614& 1& 0.620\\
spiral & 1.543& 2& 0.648\\
spiral & 1.536& 4& 0.651\\
spiral & 1.525& 3& 0.656\\
 double scroll & 1.435& 6& 0.697\\
 double scroll & 1.418&  5&  0.705\\
 double scroll & 1.396&  4&  0.716\\
 double scroll & 1.390&  L 3,R 4&  0.719\\
 double scroll & 1.368&  3&  0.731\\
 double scroll & 1.355&  L 3,R 2&  0.738\\
 double scroll & 1.346&  2&  0.743\\
 double scroll & 1.331&  2&  0.751\\
 double scroll & 1.322&  L 2,R 1&  0.756\\
 double scroll & 1.315&  L 1,R 2&  0.761\\
 double scroll & 1.314&  L 2,R 1&  0.761\\
\noalign{\smallskip}
\hline
\end{tabular}
\caption{An example of the transition from spiral to double scroll oscillation pattern of the Chua's circuit.}\label{tab:5}
\end{table}

\vskip 0.2 true cm
\leftline{{\bf A2.} The entropy produced in the $G=0.7052$ system}
\vskip 0.2 true cm
The entropy of $G=0.7052$, 7 time series simulation data are shown in Table \ref{tab:6}.  
The symbols $S_{C1}, S_{C2}, S_L$ are the zero mode subtracted Fourier amplitude squared of the capacitor 1, capacitor 2 and the inductor. The symbols, e.g. 1a, 1b, 1c correspond to the period between the pulses triggered in the sequence 1. 
\begin{table}[htb]
\begin{tabular}{llllll}
sequence& period of t&  length of the period &  $S_{C1}$&  $S_{C2}$&  $S_L$\\
\hline\noalign{\smallskip}
  1&  19.44-38.51&  1908&  1.8617&  0.02805&  2.39917\\
  \qquad 1a&  19.44-24.19&  476&  0.2067&  0.01916&  0.42803\\
  \qquad 1b&  24.20-26.40&  221&  0.39587&  0.05309&  1.50065\\
  \qquad 1c&  26.41-38.51&  1211&  0.2522&  0.02681&  0.57082\\
  2&  38.52-59.79&  2128&  1.78011&  0.02951&  2.34956\\
  \qquad 2a&  38.52-43.34&  483&  0.1635&  0.01399&  0.2815\\
  \qquad 2b&  43.35-45.29&  195&  0.22624&  0.04507&  1.12899\\
  \qquad 2c&  45.30-59.79&  1450&  0.33889&  0.03246&  0.82155\\
  3&  59.80-76.57&  1678&  1.65727&  0.02843&  2.19666\\
  \qquad 3a&  59.80-62.64&  285&  0.26386&  0.01654&  0.39284\\
  \qquad 3b&  62.65-64.60&  196&  0.233&  0.0455&  1.14973\\
  \qquad 3c&  64.61-76.57&  1197&  0.2843&  0.02828&  0.63801\\
  4&  76.58-101.76&  2519&  1.853&  0.02753&  2.37791\\
  \qquad 4a&  76.58-79.48&  291&  0.23456&  0.01512&  0.32095\\
  \qquad 4b&  79.49-81.33&  185&  0.16841&  0.03863&  0.93344\\
  \qquad 4c&  81.34-87.58&  625&  0.56976&  0.02359&  1.02295\\
  \qquad 4d&  87.59-89.84&  226&  0.41493&  0.05281&  1.50896\\
  \qquad 4e&  89.85-101.76&  1192&  0.2547&  0.02589&  0.55382\\
  5&  101.77-125.97&  2421&  1.64963&  0.03277&  2.29042\\
  \qquad 5a&  101.77-106.52&  476&  0.20028&  0.01854&  0.40951\\
  \qquad 5b&  106.53-108.69&  217&  0.37711&  0.05305&  1.48203\\
  \qquad 5c&  108.70-125.97&  1728&  0.34053&  0.03405&  0.88238\\
  6&  125.98-160.12&  3415&  1.8443&  0.02943&  2.41251\\
  \qquad 6a&  125.98-128.89&  292&  0.23172&  0.01499&  0.31587\\
  \qquad 6b&  128.90-130.72&  183&  0.16209&  0.03752&  0.91628\\
  \qquad 6c&  130.73-138.38&  766&  0.42249&  0.01906&  0.79026\\
  \qquad 6d&  138.39-140.45&  207&  0.30744&  0.05128&  1.34918\\
  \qquad 6e&  140.46-160.12&  1967&  0.35706&  0.03242&  0.88141\\
  7&  160.13-177.61&  1749&  1.55823&  0.0279&  2.0895\\
  \qquad 7a&  160.13-163.46&  334&  0.37432&  0.01719&  0.65062\\
  \qquad 7b&  163.47-165.48&  202&  0.27258&  0.04889&  1.26092\\
  \qquad 7c&  165.49-177.61&  1213&  0.25457&  0.02718&  0.58073\\
\noalign{\smallskip}
\hline
\end{tabular}
\caption{The entropy of various oscillation pattern of $G=0.7052$ samples}\label{tab:6}
\end{table}

\vskip 0.2 true cm
\leftline{{\bf A3.} The entropy produced in the $G=0.7429$ system}
\vskip 0.2 true cm
The entropy of $G=0.7429$, 9 time series simulation data are shown in Table \ref{tab:7}.  
The symbols $S_{C1}, S_{C2}, S_L$ are the zero mode subtracted Fourier amplitude squared of the capacitor 1, capacitor 2 and the inductor. The symbols, e.g. 1a, 1b, 1c correspond to the period between the pulses triggered in the sequence 1.
\begin{table}[htb]
\begin{tabular}{llllll}
 sequence&  period of t&  length of the period&  $S_{C1}$&  $S_{C2}$&  $S_L$\\
\hline\noalign{\smallskip}
 1&  19.25-32.54&  1329&  1.24631&  0.01487&  1.50291\\
 \qquad 1a&  19.25-22.19&  294&  0.14618&  0.006&  0.16314\\
 \qquad  1b&  22.20-24.26&  207&  0.09467&  0.01756&  0.42977\\
 \qquad  1c&  24.27-32.54&  828&  0.02763&  0.01711&  0.4507\\
  2&  32.55-53.68&  2114&  1.00221&  0.01189&  1.20997\\
  \qquad 2a&  32.55-35.68&  309&  0.18756&  0.00814&  0.26958\\
  \qquad 2b&  35.69-38.10&  247&  0.21155&  0.02241&  0.65649\\
  \qquad 2c&  38.11-53.68&  1558&  0.15711&  0.01091&  0.2747\\
  3&  53.69-67.16&  1347&  1.22&  0.01578&  1.49536\\
  \qquad 3a&  53.69-56.62&  293&  0.1428&  0.00587&  0.15884\\
  \qquad 3b&  56.63-58.68&  206&  0.09219&  0.01729&  0.42278\\
  \qquad 3c&  58.69-67.16&  848&  0.28869&  0.01861&  0.49672\\
  4&  67.17-89.08&  2192&  1.0097&  0.00992&  1.17949\\
  \qquad 4a&  67.17-70.10&  294&  0.15461&  0.00625&  0.18185\\
  \qquad 4b&  70.11-72.25&  215&  0.11608&  0.0193&  0.48424\\
  \qquad 4c&  72.26-89.08&  1683&  0.14373&  0.00926&  0.2377\\
  5&  89.09-105.61&  1653&  1.15233&  0.01411&  1.39947\\
  \qquad 5a&  89.09-92.08&  300&  0.17417&  0.00694&  0.23253\\
  \qquad 5b&  92.09-94.43&  235&  0.17418&  0.02234&  0.60252\\
  \qquad 5c&  94.44-105.61&  1118&  0.19562&  0.01412&  0.34302\\
  6&  105.62-133.28&  2767&  0.85158&  0.01427&  1.10939\\
  \qquad 6a&  105.62-108.66&  305&  0.18279&  0.00724&  0.25467\\
  \qquad 6b&  108.67-111.08&  242&  0.19716&  0.02276&  0.63776\\
  \qquad 6c&  111.09-133.28&  2220&  0.23936&  0.01424&  0.45321\\
  7&  133.29-150.71&  1743&  1.06588&  0.01376&  1.30729\\
  \qquad 7a&  133.29-137.17&  389&  0.24625&  0.00644&  0.3621\\
  \qquad 7b&  137.18-139.52&  235&  0.17805&  0.02235&  0.61051\\
  \qquad 7c&  139.53-150.71&  1119&  0.19759&  0.01435&  0.34891\\
  8&  150.72-167.21&  1650&  1.15656&  0.01408&  1.40273\\
  \qquad 8a&  150.72-153.70&  299&  0.17213&  0.00687&  0.22618\\
  \qquad 8b&  153.70-156.02&  232&  0.16718&  0.022&  0.59081\\
  \qquad 8c&  156.02-167.21&  1119&  0.19775&  0.0142&  0.34696\\
  9&  167.22-197.70&  3049&  0.80684&  0.01508&  1.07902\\
  \qquad 9a&  167.22-170.25&  304&  0.18018&  0.0072&  0.24926\\
  \qquad 9b&  170.26-172.65&  240&  0.1925&  0.02269&  0.63215\\
  \qquad 9c&  172.66-197.70&  2505&  0.24864&  0.01524&  0.48473\\
\noalign{\smallskip}\hline
\end{tabular}
\caption{The entropy of various oscillation patterns of $G=0.7429$ samples.}\label{tab:7}
\end{table}



\end{document}